\documentclass[%
reprint,
superscriptaddress,
nofootinbib,
amsmath,amssymb,
pra,
]{revtex4-1}
\usepackage{}\pdfoutput=1

\usepackage{graphicx}

\usepackage{bm}
\usepackage{hyperref}

\usepackage[utf8]{inputenc}
\usepackage{graphicx}
\usepackage{bm}
\usepackage{hyperref}

\usepackage[utf8]{inputenc}

\usepackage{amsmath,amssymb}

\usepackage[T1]{fontenc} 
\usepackage{microtype} 
\usepackage[english]{babel} 
\usepackage{booktabs} 
\usepackage[utf8]{inputenc}
\usepackage{amsmath}
\usepackage{graphicx}
\usepackage{amssymb}
\usepackage{braket,mleftright}
\usepackage{empheq}
\usepackage{subfigure}
\usepackage{stackrel}
\usepackage{soul}
\usepackage{blkarray}
\usepackage{multirow}
\usepackage{amsmath}
\usepackage{physics}
\usepackage{amsfonts}
\usepackage{bm}
\usepackage{bbold} 
\usepackage{color}

\usepackage[utf8]{inputenc} 
\usepackage{amsmath}

\usepackage{enumitem} 
\setlist[itemize]{noitemsep} 

\usepackage{hyperref} 
\usepackage{braket}
\usepackage{verbatim} 

\usepackage{float}
\makeatletter
\let\newfloat\newfloat@ltx
\makeatother
\usepackage{algorithm}
\usepackage[noend]{algpseudocode}
\def\ni{\noindent}

\def\be{\begin{equation}}
\def\ee{\end{equation}}

\def\bs{\begin{split}}
\def\es{\end{split}}

\def\ba{\begin{eqnarray}}
\def\bea{\begin{eqnarray}}

\def\tea{\end{eqnarray}}
\def\ea{\end{eqnarray}}
\def\eea{\end{eqnarray}}
\def\w{\omega}

\def\D{\Delta t}
\def\t{\tau}

\def\s{\sigma}
\def\p{\partial}
\def\eye{\mathbb{1}}

\def\dag{^{\dagger{}}}

\def\ep{\epsilon}

\usepackage{bm}

\def\kp{\ket{\psi}}

\def\kpt{\ket{\psi(t)}}

\def\kvC{\ket{v_0}}

\def\da{\downarrow}
\def\ua{\uparrow}

\usepackage{mathtools}

\def\R{\mathbb{R}}
\usepackage{amssymb}

\def\p{\partial}
\def\w{\omega}
\def\dag{^{\dagger{}}}
\def\be{\begin{equation}}
\def\ee{\end{equation}}

\def\bs{\begin{split}}
\def\es{\end{split}}

\def\L{\mathcal{L}}
\def\KN{\mathcal{K}_{N}}

\def\R{\mathcal{R}}
\def\Rm{\Tilde{\mathcal{R}}}

\begin{document}

\title{K-GRAPE: A Krylov Subspace approach for the efficient control of quantum many-body dynamics}

\author{Mart\'{i}n Larocca}
\email{mail to: larocca@df.uba.ar}

\affiliation{Departamento de F\'{i}sica “J. J. Giambiagi” and IFIBA, FCEyN, Universidad de Buenos Aires, 1428 Buenos Aires, Argentina
}
\author{Diego Wisniacki}

\affiliation{Departamento de F\'{i}sica “J. J. Giambiagi” and IFIBA, FCEyN, Universidad de Buenos Aires, 1428 Buenos Aires, Argentina
}%

\date{October, 2020}%

\begin{abstract}




The Gradient Ascent Pulse Engineering (GRAPE) is a celebrated control algorithm with excellent converging rates, owing to a piece-wise-constant ansatz for the control function that allows for cheap objective gradients. However, the computational effort involved in the exact simulation of quantum dynamics quickly becomes a bottleneck limiting the control of large systems. In this paper, we propose a modified version of GRAPE that uses Krylov approximations to deal efficiently with high-dimensional state spaces. Even though the number of parameters required by an arbitrary control task scales linearly with the dimension of the system, we find a constant elementary computational effort (the effort per parameter). Since the elementary effort of GRAPE is super-quadratic, this speed up allows us to reach dimensions far beyond.
The performance of the K-GRAPE algorithm is benchmarked in the paradigmatic XXZ spin-chain model.

\end{abstract}

\maketitle

\section{Introduction} \label{Section-Intro}
At the core of quantum technology is our ability to control quantum dynamics. In the last decades, we have evidenced unprecedented advances in the manipulation of dynamical processes at the atomic and molecular scale. The control is usually enforced by applying properly tailored external electromagnetic fields. One prosperous framework for producing these control fields is Quantum Optimal Control (QOC) \cite{dalessandro,Brif_2010,glaser2015training}. QOC methods have thrived at a range of emerging quantum technologies, e.g. communication, computation, simulation and sensing —until now, at the level of a few qubits \cite{comparing,wilhelm2013transmon,nitrogen,machnes2020integrated,qoc_supercond2020}.





A notable member of QOC's "zoo" of algorithms is the Gradient Ascent Pulse Engineering (GRAPE), first introduced in the context of NMR spectroscopy \cite{bib:grape1}. As its name suggests, it proposed a gradient-based optimization of the control protocols, as opposed to the derivative-free (finite-difference) approaches that were commonly used at the time. The key to GRAPE's success was to propose a piece-wise-constant (PWC) ansatz for the control that in turn allowed for cheap gradients of the objective. Gradient based algorithms usually have much better convergence than the gradient-free \cite{nocedal2006numerical}. Its ability to produce high-quality optimal controls in an inexpensive and fast fashion made it the state of the art algorithm in quantum control. 

Despite its successful application in small systems, QOC methods encounters severe limits when applied to many-body quantum systems. Due to the exponential complexity of simulating the latter, control algorithms fail to yield a desired final state within an acceptable computational time. There are basically two approaches for the efficient simulation of quantum evolution. A first kind assumes that entanglement will be small during the whole evolution and uses a truncated representation of the state vector.  This is the realm of tensor network methods, such as the density-matrix renormalization group
evolution or tDMRG \cite{scholl1,scholl2}. A second alternative are Krylov subspace methods. These circumvent the computationally impracticable task of diagonalizing the full Hamiltonian, a key step in the computation of matrix exponentials arising in the treatment of time evolution, by considering only a reduced number of effective energy levels \cite{hock,time-evo}. Just to illustrate the power of the method, Ref. \cite{massive} reports having been able to simulate time-evolution in Hilbert subspaces of dimensions up to 9 billion, using parallel supercomputers.

A number of proposals for controlling many-body dynamics following the first approach were made \cite{doria,calarc2016scirep,sherson2020achieving}. Refs. \cite{doria,calarc2016scirep} used a matrix-product-state ansatz with a derivative-free approach to drive a superfluid-Mott insulator transition in an optical lattice. The latter \cite{sherson2020achieving}, revisited the problem using gradient based optimization and achieved much better fidelities. Let us note that this task, the connection of ground states on both sides of a phase transition, is perfectly suited for such low-entanglement ansatze. Nevertheless, more general control scenarios may require full state descriptions.

In this paper, we explore the possibility of using the second alternative in a control context. We present a modified GRAPE algorithm that uses Krylov approximations instead of full Hamiltonian eigendecompositions. As a particular example, we try to control pure-state transitions on an XXZ spin-chain model. Fixing the number of effective levels, we are able to locate optimal protocols with a computational effort per parameter roughly independent of the size of the system. Its dimension only affects the search effort through the number of parameters that are required for control solutions to exist, a quantity that grows linearly with such. In an $D$-dimensional Hilbert space, at least $2D-2$ parameters are needed to control pure-state transfers \cite{natural,beltrani,moore2012exploring}. This \textit{minimum number of parameters} is related to the informational content of the control field (basically $2D-2$ real numbers are needed to specify an arbitrary pure state of an N dimensional system). Because our method uses the entire representation of quantum states, we are able to manipulate arbitrary entangled states.

The paper is organized as follows. In Section \ref{sec-2}, we introduce K-GRAPE algorithm, a modification of GRAPE, with time evolution 
approximated by Krylov subspace methods. In Sec.  \ref{sec-3}, the algorithm is carefully tested to control a XXZ spin-chain model.
In this Section we describe the control task and the numerical results. Finally, in Section \ref{Section-con}, we draw conclusions on the results obtained. To make the paper self-contained, we have a included Appendices on Krylov's approximation and on the GRAPE algorithm.


\section{The K-GRAPE algorithm}\label{sec-2}

\begin{figure}
	\begin{center}
		\includegraphics[width=.4\textwidth]{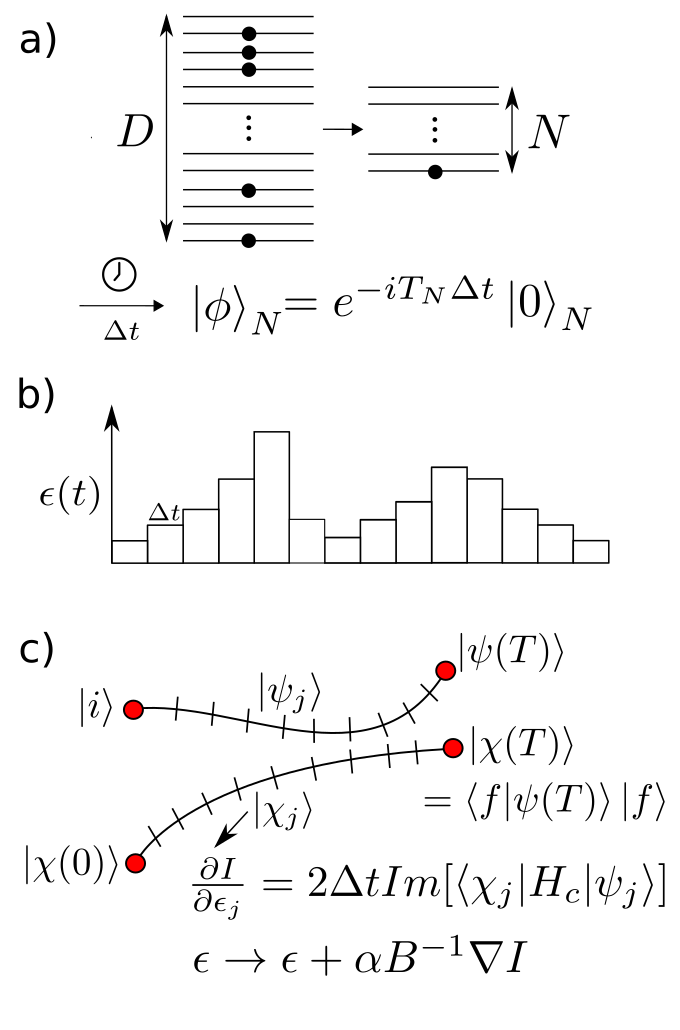}
		\caption{Artist's impression. (a) Krylov's approximation: an initial state with an arbitrary distribution on the full D-dimensional state space is mapped into the ground state of an effective $N$-level system. This ground state evolves under the action of $T_N$, the Hamiltonian in the reduced Krylov basis. Finally, the evolved state on the full Hilbert space is recovered with the inverse mapping. (b) GRAPE's ansatz: a piece-wise-constant control field. (c) the zeroth order approximation of the gradient is equivalent to the bracket of the control Hamiltonian between the forwards evolved initial state and the backwards evolved target-projected final state at each time-slot. At each iteration, the control field $\epsilon$ is updated using the gradient $\nabla I$, the inverse of an approximated Hessian $B$ and a backtracked step length $\alpha$. }
		\label{f0}
	\end{center}
\end{figure}


Krylov subspace methods are well known linear algebra techniques usually used to approximate the action of the function of a matrix on a vector \cite{hock,time-evo}. The Krylov subspace $\KN$ of a Hamiltonian $H$ and a state $\kp$ is defined as the span of $\{ \kp, H\kp,\dots,H^{N-1} \kp\}$. In the following we will use Krylov subspaces of dimension $N\ll D$, but let us stress that the case $N=D$ (the so called Krylov space) not necessarily corresponds to the Hilbert space of the system. Only systems with a non-degenerate energy spectrum explore the entire state space and thus their Krylov spaces have dimension $D$. An orthonormal basis spanning $\KN$ can be used to perform extremely cheap and accurate approximations to the time-evolution of a state. The basic mechanism first maps the initial state into the ground state of an effective $N$-level system (see Fig. \ref{f0}.(a)). Time-evolution is computed in this reduced space (and thus very efficiently) using $T_N$, the projection of the Hamiltonian in the Krylov basis $\ket{\phi}_N = e^{-i T_N \D}\ket{0}_N$. Finally, the map is inverted: the actual "full-dimensional" evolved state is recovered as a linear combination of the Krylov basis vectors, using the amplitudes of the subspace-evolved $\ket{\phi}_N$ as coefficients. For details on this procedure, see Appendix \ref{ap-kry}. We will proceed to introduce the control scenario.

 Consider a typical controlled system, where the Hamiltonian,

\be
H(t) = H_d+\ep(t)H_c,
\label{htot}
\ee

\ni is tunable through the time-dependent control $\ep(t)$. Here, $H_d$ and $H_c$ are usually addressed as the \textit{drift} and \textit{control} terms, respectively, and their nested commutators determine the degree of \textit{controllability} of the system \cite{dalessandro}. In short, how much of Hilbert space can be dynamically explored by arbitrary choices of $\ep(t)$. The shaping of this function, also referred to as the \textit{protocol} or \textit{control field} (experimentally, control is usually enforced through dipole couplings with electromagnetic fields), will allow us to govern the evolution of the system.  For example, consider the situation where the system is initially in a given state $\ket{i}$ and we are interested in a dynamics that prepares the target state $\ket{f}$ at time $t=T$. In order to search for controls that accomplish the task, one has to introduce a figure of merit quantifying the degree of fulfilment. In terms of the overlap $\beta=\bra{f} U(T) \ket{i}=\braket{f}{\psi(T)}$, we define the infidelity

\be
I =1- |\beta|^2
\label{objective}.
\ee

\ni Here (and throughout) we have avoided stating explicitly the dependence of $I$, $\beta$ and $U$ on the protocol $\ep(t)$, given by the fact that the latter is the solution of

\begin{equation}
\begin{split}
&i\frac{dU(t)}{dt}=H[\ep(t)]U(t) \\
&U(0) = \eye
\end{split}
\label{sch}
\end{equation}

\ni evaluated at $t=T$. The functional of Eq. (\ref{objective}) is a map from the space of protocols to the interval $[0,1]$ of real numbers. This map is usually known as the Quantum Control Landscape \cite{rabitz_qcl,larocca2019exploiting}. Solutions to the control problem are global minima of the landscape. Thus, the problem of finding suitable controls is equal to a minimization process in this infinite-dimensional space. In practice, optimal control techniques introduce a parametrization on the field and thus constrain the problem to finite-dimensional search spaces. For example, GRAPE algorithm uses a PWC parametrization of the control, namely,

\begin{equation}
    \ep(t)=
    \begin{cases}
        \ep_1 & \text{if $0<t<\Delta t$}\\
        \vdots \\
        \ep_M & \text{if $(M-1)\Delta t<t<T$}.\\
    \end{cases}
\label{PWC}
\end{equation}

\ni Here, the protocol duration was divided in $M$ uniform intervals $\D=T/M$ at which the control is constant (see Fig. \ref{f0}.(b)). This ansatz has several advantages. First, the propagator factorizes into a product of individual sub-propagators, each of which is generated by a constant Hamiltonian and thus has a simple matrix exponential form. More importantly, the derivatives of the objective with respect to the controls are also extremely simple, in particular to first order in $\D$ (see Appendix \ref{ap-grape} for a detailed derivation),

\be
\frac{\p I}{\p \ep_j} \approx   -2 \D \Im[\bra{\chi_j} H_c \ket{\psi_j}]
\label{gradI}
\ee

\ni where $\ket{\chi_j}=\beta U_{j+1}\dag \ldots U_M \dag \ket{f}$ and $\ket{\psi_j}=U_j\ldots U_1\ket{i}$ can be pictured as forward and backwards propagated states. More precisely, $\ket{\chi_j}$ is the final time evolved state $\ket{\psi(T)}=U\ket{i}=\ket{\psi_M}$, projected into the target state and backwards evolved (see Fig. \ref{f0}.(c)). A sketch of GRAPE's framework is presented in Algorithm \ref{alg-GRAPE}. A general update rule uses a positive definite matrix B and the gradient $\nabla I$ to generate a search direction $p=B^{-1}\nabla I$ and then takes an appropriate step $\alpha$ in that direction. In the simplest situation (steepest descent) $B$ is the identity matrix and thus the step is taken in the direction of the gradient. Newton's method instead uses the Hessian:  $B=\nabla^2I$. There are two problems with this: (i) computing the Hessian is expensive, and (ii) the Hessian may not be positive definite and thus have no inverse. The BFGS Method \cite{BFGSDavidon1959} uses the gradient to build very cheap approximations of the Hessian (the approximation is not built entirely at each iteration, but instead only updated where relevant) that in turn are positive definite. Modern implementations of GRAPE use the L-BFGS method \cite{nocedalLBFGS}, a limited-memory variant that achieves super-linear convergence in a time and memory efficient manner. At each iteration, the step $\alpha$ is chosen with a backtracking routine such that it satisfies Wolfe's conditions \cite{nocedal2006numerical}.

    \begin{algorithm}
        \caption{Basic GRAPE algorithm}
        \label{alg-GRAPE}
        \begin{algorithmic}[1]
            \State $\text{Guess initial control amplitudes } \{\ep_j\}$
            \While{$\text{not converged}$}
            \State $\text{compute propagators } U_j \gets e^{-i H_j \D}$
            \State $\text{forwards evolve } \ket{\psi_j} \gets U_j \ket{\psi_{j-1}}$
            \State $\text{set } \ket{\chi_M} \gets \braket{f}{\psi_M} \ket{f}$
            \State $\text{backwards evolve } \ket{\chi_j} \gets U_j \dag \ket{\chi_{j+1}}$
            \State $\text{evaluate gradient } \frac{\p I}{\p \ep_j} \approx   -2 \D \Im[\bra{\chi_j} H_c \ket{\psi_j}]$
            \State $\text{update amplitudes } \ep \gets \ep+\alpha B^{-1}\nabla I \text{ and go to 3}$
            \EndWhile
        \end{algorithmic}
    \end{algorithm}

GRAPE quickly became widely used in the control community. It offered the possibility of optimizing control pulses in much larger control spaces than those allowed by finite-difference gradients. Note that these require at least $M+1$ full time evolutions (in the forward-difference setting), as compared with the two evolutions required by GRAPE. Over the years, several enhancements to the method were proposed \cite{ciama2011,defoquieres2011_secondorder,comparing,goodwin}. In particular, it was realized that the eigenvalues and eigenvectors of each of the constant Hamiltonians, used to compute the matrix exponentials for the objective, could be cached and re-used to compute exact gradients \cite{comparing}. This setting is faster, but for large system dimensions it quickly becomes impractical in terms of memory, considering we have to store $M$ (one for each parameter) D-dimensional matrices  and that the number of parameters scales at least linearly with D. In short, it requires storing more and bigger matrices.



The main result of this work is that the GRAPE algorithm can be slightly modified to perform efficiently in the near many-body regime. Instead of computing exact forward and backwards state propagations, we propose to use Krylov approximations. We use a centered version of the zeroth order gradient approximation (Eq. (\ref{gradI})) which we feed to a L-BFGS routine. We note that although there's plenty of room for playing with more accurate higher order approximations (see Eq. (\ref{gradp})) or the numerical integration of the natural gradient ( Eq. (\ref{natgrad})), we found the simple centered zeroth order gradient more than enough for a first demonstration of the virtues of the proposed method. In the following, we define a particular control setting and compare the performance of GRAPE and K-GRAPE algorithms. We use a built-in implementation of GRAPE offered by the python toolbox QuTiP \cite{bib:qutip,bib:qutip2}, that is based on the DYNAMO package \cite{comparing}.  In order to be able to reach large dimensions, we chose not to cache the propagator gradients and instead let the fidelity computer calculate them as needed.

\section{Example: controlling a XXZ Heisenberg spin chain}  \label{sec-3}
In this section, we test the proposed K-GRAPE algorithm in the well-known XXZ spin-chain. We benchmark the new procedure comparing 
its performance with the usual GRAPE algorithm.

\subsection{Spin chain model and control task} \label{sec-spin}

Let us consider a one-dimensional system consisting of L spin $1/2$ particles  whose drift evolution is described by the XXZ Hamiltonian \cite{poggichaos2016},

\be
H_d = \frac{J}{2} \sum_{i=1}^{L-1} \sigma_{i}^{x} \sigma_{i+1}^{x}+\sigma_{i}^{y} \sigma_{i+1}^{y}+\alpha_{z} \sigma_{i}^{z} \sigma_{i+1}^{z}.
\label{H01}
\ee

\ni Here, $\sigma^{x,y,z}_i$ are the Pauli matrices   for the ith particle, and we have set $\hbar=1$ such that energy is measured in units of the interaction strength $J$ and time is measured in units of $J^{-1}$. This Hamiltonian has a number of symmetries. First, it conserves the total magnetization in the z direction, $\sigma_{z}=\sum_{i=1}^{L} \sigma_{i}^{z}$, since $[H_d,s_z]=0$. This allows us to fracture the entire state space into subspaces $S_K$ of fixed number of excitations

\be 
\mathcal{H}=\bigoplus_{K=0}^{L} \mathcal{S}_{K}
\ee

\ni The dimension of this subspaces is simply the number of ways of picking $K$ excitations out of L sites, and is given by

\be
D_K = \binom{L}{K} = \frac{L!}{K!(L-K)!}.
\ee

\ni Additionally, the Hamiltonian in Eq. (\ref{H01}) conserves parity. The parity operator acts on a given computational state by "mirroring" against the middle of the chain, e.g.

\be
\Pi \ket{\da \da \ua \da} = \ket{\da \ua \da \da} .
\ee

\ni Because the couplings in $H_d$ are homogeneous, this two operators commute $[H_d,\Pi]=0$. Thus, each excitation subspace $S_k$ is further broken into two parity-excitation (PE) subspaces, $S_{k,+}$ and $S_{k-}$, with even and odd parity, respectively.  Finally, we will avoid conservation of $S^2$ by choosing $\alpha_z = 0.5$ and set the coupling strength to $J = 1$. 

As a control Hamiltonian, we use
\be
H_c = \frac{J}{2}( \s^z_1 + \s^z_L)
\ee
\ni such that the total Hamiltonian $H(t)$ in Eq. (\ref{htot}) still commutes with $\sigma_z$ and $\Pi$ for any choice of control function. Initial eigenstates of these operators evolve constrained to the PE subspaces, thus allowing us to use a reduced representation in the computations. The Hamiltonian in the reduced PE subspace is constructed in the following way: (i) write the parity operator in a given excitation subspace $S_K$ and (ii) diagonalize it, (iii) use the eigenvectors associated with the desired parity (we use even parity throughout) to build a rectangular change of basis matrix Q and (iv) reduce the Hamiltonian, $H_{red}=QH Q\dag $. The computational basis vectors of the reduced excitation subspace are ordered relative to the binary number associated with each sequence of bits (considering spin ups as zeros and spin downs as ones), in ascending order.


As a control task, we will attempt to drive the first coordinate vector in a given PE subspace into the last one 

\be
e_1 \xrightarrow[\ep(t)]{} e_D.
\label{PST}
\ee

\ni In the particular case of an odd number of spins and an even number of excitations, this task corresponds to the transport of a cluster of excitations from the middle of the chain to the edges, for example

\be
\begin{split}
    e_1 &= \ket{\da \cdots\cdots \da \ua \cdots \ua \da \cdots\cdots \da} \\
    e_D &= \frac{1}{\sqrt{2}}(\ket{\ua \cdots \ua \da \cdots\cdots \da}+\ket{\da \cdots\cdots \da \ua \cdots \ua}) .
\end{split}
\ee

 Let us mention that we've performed numerical controllability tests (using the nested commutators of the reduced $H_0$ and $H_c$, as described in \cite{dalessandro}) and found the PE subspaces to be controllable.

\subsection{Numerical results}\label{sec-num}

In the following we present a numerical study of the performance of K-GRAPE on the spin-chain pure-state transfer task defined above, using different choices of length $L$ and excitations $K$. The results are properly compared with the control using the GRAPE algorithm. We initialize random fields with $M=4D$ parameters drawn from a uniform distribution in $[-1,1]$. This linear scaling of $M$ with dimension is roughly twice the strictly needed\footnote{There's a lower bound on the minimum number of parameters for solutions to arbitrary pure-state transfer problems to exist \cite{moore2012exploring}}. We further fix $\D=.5$ and truncate at $N=10$ Krylov vectors. A target infidelity of $I_{target} = 10^{-2}$ is set and as an additional stopping criterion, the minimum change in the objective from one iteration to the
next is chosen to be $\Delta I_{min} = 10EPS$, with EPS the machine precision. In the following we present run-time data, which was measured in seconds \footnote{The optimisations were carried out on an Intel(R) Core(TM) i7-8550U CPU @ 1.80GHz with 16GB of RAM.}.



\begin{figure}
	\begin{center}
		\includegraphics[width=.5\textwidth]{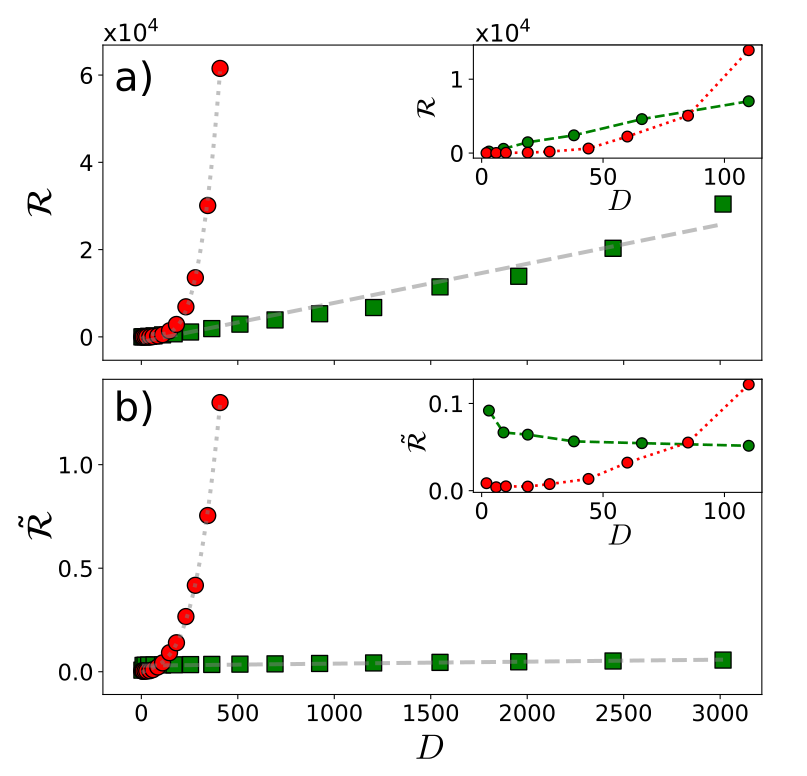}
		\caption{Algorithm benchmarking: (a) optimization run-time $\R$ and (b) elementary run-time $\Tilde{\R}$ (run-time per iteration per time-step) for the state transfer task in Eq. (\ref{PST}) using GRAPE (dotted line and red squares) and K-GRAPE (dashed line and green squares) as a function of D, the dimension of the subspace holding the dynamics. The data points corresponding to GRAPE and K-GRAPE are fitted with linear and a cubic functions, respectively. In a first low-dimensional regime GRAPE outperforms (see inset), while in the large-dimensional one, K-GRAPE does.
		}
		\label{dip}
	\end{center}
\end{figure}

Fig. \ref{dip}.(a) shows the total run-time $\R$ consumed by GRAPE (dotted line and red squares) and K-GRAPE (dashed line and green squares) as a function of dimension D. Each data point corresponds to a single successful optimization. We note that almost no traps were observed with these settings. The results correspond to chains with $K=3$ excitations in the case of GRAPE and $K=4$ in the case of K-GRAPE. Other values of $K$ were tested and were found to have no influence on the results. On a first low-dimensional regime, GRAPE outplays (see inset). Instead, for $D\geq100$, K-GRAPE is clearly more efficient \footnote{If the caching of propagator gradients is enabled, GRAPE performs much faster but still super-quadratically, meaning the intersection with K-GRAPE is only pushed forwards. We built a run-time curve (similar to those of Fig. (\ref{dip})) up to $D=255$ (were the routine collapsed due to memory overloading), fitted the data and estimated the crossing to be at $D=400$.}. The gray lines connecting the data are linear and cubic fits, for K-GRAPE and GRAPE respectively. Fig. \ref{dip}.(b) provides the "elementary" run-time $\Rm$, defined as the run-time per field evaluation (iteration) and per time-step. This elementary run-time is independent of dimension for K-GRAPE and is at least quadratic for GRAPE. This is reasonable since K-GRAPE uses a fixed size effective Hamiltonian while GRAPE has to deal with D-dimensional matrices. GRAPE's effort is well explained in terms of the eigendecompositions at its core \cite{scaling}. In turn, this elementary efforts account for the total run-times observed. Since both algorithms perform a number of iterations that is roughly independent of D, the only dependence of the full run-time with dimension is through the number of parameters, that must grow linearly with dimension to satisfy the control constraint. Thus, for example,  K-GRAPE's constant elementary effort is translated into a linear scaling. Let us note that we have checked that the error between exact and Krylov propagated final states is orders of magnitude below the target infidelity.

\begin{figure}
	\begin{center}
		\includegraphics[width=.5\textwidth]{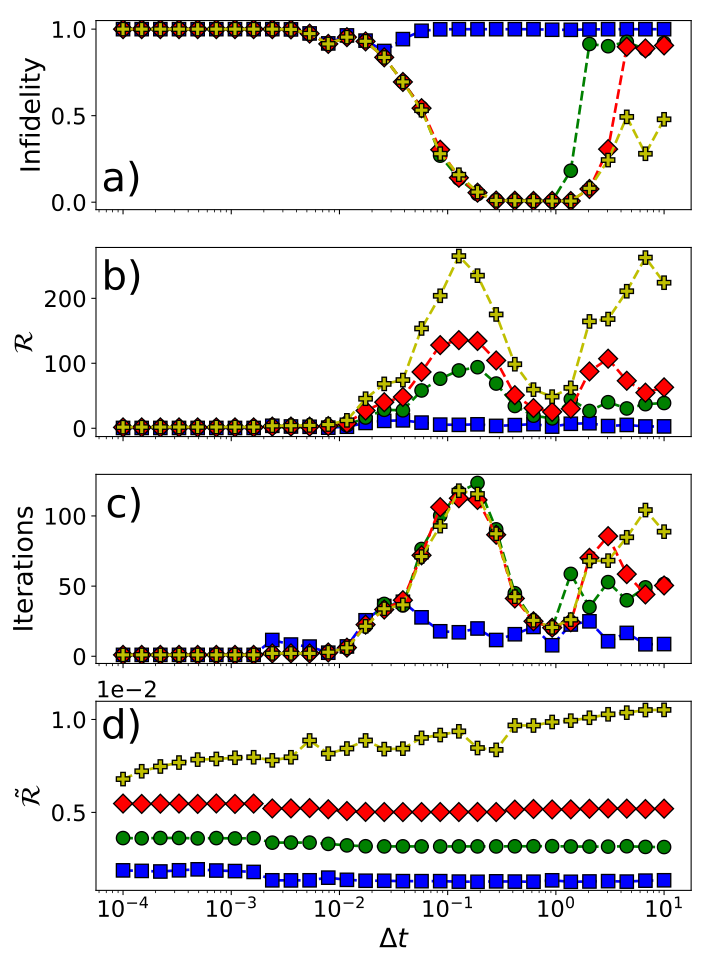}
		\caption{ Time-step study: (a) minimum infidelity, (b) mean run-time $\R$, (c) mean iterations and (d) $\Tilde{\R}$ run-time per iteration per time step, as a function of $\D$. The different curves correspond to different values of the truncation parameter, $N=2,10$ and $18$, marked with blue squares, green circles and red diamonds, respectively. The yellow pluses correspond to the centered zeroth order gradient, evaluated using an exact evolution. See text for details. }
		\label{vsdt}
	\end{center}
\end{figure}

The performance of K-GRAPE depends critically on the choice of time step $\D$. On one hand, the algorithm is built upon an approximated gradient that works optimally in the low $\D$ regime. On the other, there is a minimum time $T_{min}$ (lower bounded by the quantum speed limit time) such that the control problem has solutions. That is, with a too-small value of $\D$ (and a fixed number of parameters) the algorithm will converge properly but will not be able to attain good fidelities simply because they do not exist. Finally, the quality of the gradient is further tied to the quality of the Krylov approximation, which depends on a large-enough truncation $N$ and, again, on a small-enough time-step $\D$. 

In order to study the behavior of K-GRAPE with the time step, we initialize and optimize $20$ seeds for different values of $N$ and $\D$. The number of parameters is still fixed at $M=4D$ (and will be fixed throughout) and we use $D=60$.  Fig. \ref{vsdt} (a) shows the minimum infidelity achieved as a function of $\D$ for $N=2,10$ and $18$ (blue squares, green circles and red diamonds, respectively). The yellow pluses correspond to an exact evolution (no Krylov) and a centered zeroth order gradient.  As mentioned above, there is a trade-off in $\D$: small values prohibit the state transfer while large values compromise the gradient and thus convergence. In the middle there's a "control window". Notice how too-small values of truncation (e.g. $N=2$) also damage the gradient and in consequence no window is observed. Instead, for $N=18$ we find a broader window that in the case of $N=10$. Panel (d) displays the elementary effort, which is observed to grow with $N$. The mean number of iterations, shown in panel (c), is seen to present a "bump" in the region around $\D_{min}=T_{min}/M\approx 0.25$. We think this could be related to a blossoming of traps in the control landscape \cite{laro2018}. To the right of this maximum, the iterations decrease and then grow back again, this time due to the growing inaccuracies in the gradient. Finally, these aspects merge in the total run-time observed (panel (b)). Note that the observed $\D_{min}$ corresponds to a $T_{min}\approx D$. We have numerically checked this relation to hold for a wide range of dimensions. In particular, the benchmarking study of Fig. \ref{dip} is consistent with this estimation, since we chose $\D=.5$ (twice this minimum value) and found solutions every time. We cannot give a precise explanation of why this is the case, but here is an attempt. If one considers the quantum speed limit time associated with the drift Hamiltonian (that in this case is $\tau_{qsl}\approx1$) as a bound on the time needed to reach an orthogonal state, and one assumes that, in the worst case scenario, the trajectory explores all the $D-1$ orthogonal states before reaching the desired target, we can argue that the minimum control time is bounded by $T_{qsl}\approx D-1$ \cite{Poggi_2013_QSL,calcarco_commu2011}.

\begin{figure}
	\begin{center}
		\includegraphics[width=.5\textwidth]{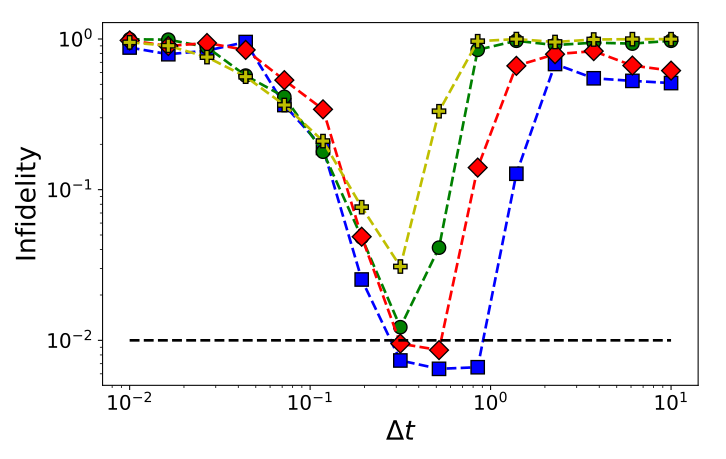}
		\caption{"Death" of a control window: minimum infidelity as a function of time-step $\D$ for a fixed truncation of $N=6$ and different dimensions $D=10,19,44$ and $146$ (blue squares, green circles, red diamonds and yellow pluses).}
		\label{fig death}
	\end{center}
\end{figure}

Let us further characterize the control windows observed in Fig. \ref{vsdt}.(a). To do so, we repeat the previous study, this time as a function of dimension (see Fig. \ref{fig death}). We plot the minimum achieved infidelity as a function of time-step $\D$ for a fixed truncation of $N=6$ and different dimensions $D=10,19,44$ and $146$ (marked with blue squares, green circles, red diamonds and yellow pluses). We find that while the truncation is enough to control systems of dimensions $D=10$ and $19$ (see how the blue and red dots do find infidelities below the target, plotted as a black dotted line), the control window "shutters" for greater dimensions. We emphasize that, due to this behaviour, we do not expect the linear behaviour of Fig. \ref{dip} to continue forever. At some large value of dimension, we expect an increase on the number of iterations (owing to the buildup of imprecisions in the gradient) followed by a loss of controllability. 

\begin{figure}
	\begin{center}
		\includegraphics[width=.5\textwidth]{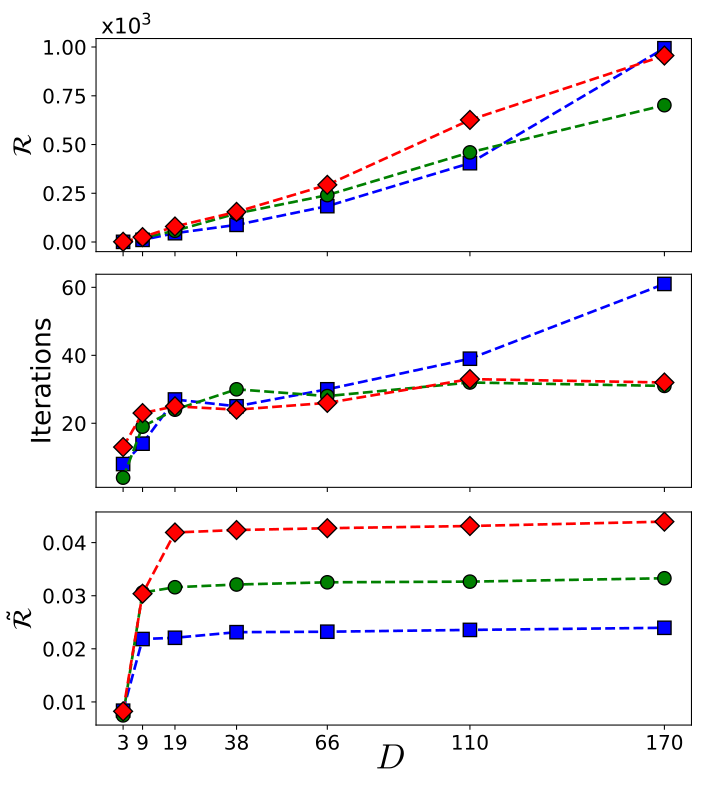}
		\caption{Truncation study: (a) run-time $\R$, (b) iterations and (c) elementary run-time $\Rm$ involved in the achievement of controls as a function of dimension D. The different curves correspond to different values of the truncation parameter $N=8,10$ and $12$, marked with blue squares, green circles and red diamonds, respectively.}
		\label{fig Nts}
	\end{center}
\end{figure}

To conclude, we present a study of the truncation parameter $N$. Fig. \ref{fig Nts} (a) shows curves of total run-time $\R$ involved in achieving an optimal control field (within the desired fidelity), as a function of dimension, for different values of truncation: $N=8,10$ and $12$, marked with blue squares, green circles and red diamonds, respectively. Since the elementary run-time $\Rm$ (panel (c)) is flat with D, and since the number of iterations (panel (b)) is roughly independent of $N$, the total run-time grows roughly linearly with $N$ (in accordance with fig. (\ref{dip}).(a)). An exception is the case of $N=8$. Here, for dimensions $D>38$, the window starts to close, the iterations grow and the total run-time becomes super-linear. To avoid this situation, either the number of parameters should be increased (such that $\D_{min}=T_{min}/M$ decreases, pulling away the lower edge of the window) or the gradient should be made more precise (kicking forwards its upper one). Either choice involves further computations, evidencing a trade-off between window width and effort.



\section{FINAL REMARKS}\label{Section-con}


Krylov subspace methods have been delivering fruitful insights and advancements in several areas of research that go from optimization theory
to the characterization of operator complexity in chaotic many-body quantum systems \cite{Gould1999SolvingTT,qucon,prxparker,2020operatorSonner}. Important for the context of this paper, several recent works have exploited its extreme efficiency at simulating time evolution on large quantum systems \cite{massive,time-evo}.


The GRAPE algorithm is an acclaimed quantum optimal control method that has enabled the efficient production of high-quality protocols to actively guide the dynamics of quantum systems. Unfortunately, the performance of GRAPE is seriously hindered when reaching out of the small dimensional regime.

In this work, we have presented an innovative control algorithm that combines these two worlds. Using truncated Krylov evolutions, we were able to efficiently power the GRAPE algorithm in the near many-body regime, where optimization becomes problematic with traditional methods. We tested K-GRAPE in a XXZ spin-chain model and demonstrated it's ability to find control solutions at dimensions far beyond the capabilities of standard GRAPE. We showed that its intrinsic complexity is independent of the dimension of the problem, as opposed to GRAPE's quadratic scaling. This speed-up allows us to control systems of dimensions clearly exceeding what was available. We want to emphasize that in no way our algorithm breaks the exponential scaling presented by many-body systems. 

To finish, we note the suitability of K-GRAPE's framework to be adapted to operator control. This straightforward generalization is very interesting since, for example, it would grant access to the efficient design of control protocols in large open quantum systems. This
investigation will be part of a forthcoming publication.

\begin{acknowledgements}
We acknowledge the valuable discussions held with P.M. Poggi. This work was partially supported by CONICET (PIP 112201 50100493CO), UBACyT (20020130100406BA) and ANPCyT (PICT-2016-1056)).
\end{acknowledgements}

\appendix

\section{The Krylov approximation}\label{ap-kry}

In this section, we describe in detail the mechanism involved in the Krylov approximation. To begin with, consider the construction of $B_N=\{\ket{v_0},\dots,\ket{v_{N-1}}\}$, an orthonormal basis spanning the Krylov subspace. This can be done using the Lanczsos method (see Algorithm \ref{lanczsos}). Notice that we only have to explicitly remove the components of $\ket{x_j}$, the new \textit{"candidate"} basis vector, on the last two elements of the basis, $\ket{v_{j-1}}$ and $\ket{v_{j-2}}$. The reason for this is that the Hamiltonian, by construction, is tridiagonal in the Krylov Basis. Moreover, the coefficients appearing in the orthonormalization procedure, $\bra{v_{j-2}}\ket{x_j}$ and $\bra{v_{j-1}}\ket{x_j}$, are the off-diagonal and diagonal entries of such tridiagonal matrix. It is important to note that if this procedure is used to build large Krylov bases, round-off errors intrinsic to floating-point arithmetic may cause loss of orthogonality between the basis vectors. An obvious way to handle this problem is to explicitly orthonormailze the new candidate against all previous vectors, as in a standard Gram-Schmidt procedure. This can become pretty expensive time and memory wise. More elaborate alternatives involve keeping track of the orthogonality loss and only perform the re-orthonormalizations when needed \cite{PRO}.

\begin{algorithm}
\caption{Lanczos Algorithm. Receives a Hamiltonian $H$ and a state $\kp$ and returns a set of $N$ orthonormal vectors $\{\ket{v_i}\}$, the Krylov basis.}\label{lanczsos}
\begin{algorithmic}[1]
\State $\kvC\gets\kp$
\For {$j >1$}
\State $\ket{x_j}\gets H\ket{v_{j-1}}$
\State $\ket{\w_j}=\ket{x_j}-\sum_{k=j-2}^{j-1} \bra{v_k} \ket{x_j} \ket{v_k}$
\State $b_j=\sqrt{\bra{\w_j}\ket{\w_j}}$
\If {$b_j>0$}
\State $\ket{v_j}\gets\frac{1}{b_j}\ket{\w_j}$.
\EndIf
\EndFor
\end{algorithmic}
\end{algorithm}


Krylov Bases have obvious application in the time-evolution of a state

\be
\begin{split}
\ket{\psi(t+\D)} &= e^{-i H \D} \kpt \\
&\approx \ket{\psi_N (t+\D)}
\end{split}
\ee

\ni Here, $\ket{\psi_N (t+\D)}\in B_N$ is a cheap yet excellent approximation of the evolved state. Let us explain how to build it. Following \cite{time-evo}, first consider the projector onto the truncated basis, $P_N = \sum_{j=0}^{N-1} \ket{v_j}\bra{v_j} = V_N\dag V_N$, where $V_N\dag$ is the $(D,N)$ change of basis matrix between the reduced $N$-dimensional Krylov basis and the original D-dimensional basis

\be
V_N\dag = \begin{bmatrix} \vdots &  \vdots  && \vdots \\ \ket{v_0}, & \ket{v_1},&  &, \ket{v_{N-1}} \\ \vdots &  \vdots  && \vdots \end{bmatrix} 
\ee

\ni The method proceeds by locating the element $\ket{\psi_N(t+\D)}\in B_N$ that is closest to the exact evolved state. This is carried out by considering the evolution with a projected propagator

\be
\begin{split}
\ket{\psi(t+\D)} &\approx P_N e^{-i H \D} P_N \kpt\\
&=V_N\dag e^{-i T_N \D} V_N \kpt
\end{split}
\ee

\ni where $T_N=V_N H V_N\dag $ is the Hamiltonian in the Krylov basis. By construction, $V_N$ maps the initial state into the ground state of an $N$-dimensional system, $V_N\kpt=(1,0,\cdots,0)^T\equiv \ket{0_N}$. This state evolves subject to $T_N$, populating these effective levels, and is finally mapped back to the full original space (see Fig. \ref{f0}).

Numerous estimations and bounds to the error in this approximation exist \cite{hock,saad,expokit}. Most of them are based under the assumption of exact arithmetic and are usually too pessimistic to explain the numerically observed error reductions. Nevertheless, we have numerically checked that in the small $\D$ regime (in particular when $\D<\frac{N^2}{W}$ with W the spectral width of the Hamiltonian) the dependence is $O(\D^N)$ \cite{time-evo}.




\section{The GRAPE Algorithm}\label{ap-grape}
Let us review the GRAPE algorithm. The PWC ansatz for the protocols (see Eq. (\ref{PWC})) induces a factorization on the propagator

\be
U(T) = U_M \cdots U_1
\ee

\ni where each of these subpropagators is generated by the constant Hamiltonian $H_j=H_0+\ep_j H_c$, correspondent with a given time slot, and thus has a simple matrix exponential form

\be
U_j = e^{-i H_j \D}
\label{eq-matexp}
\ee

\ni Moreover, the derivatives of the objective with respect to the controls are also extremely simple. First note that the gradient of the objective is related to the gradient of the propagator through

\be
\nabla I = -2\Re{\bra{f}\nabla U(T) \ket{i} \bra{i} U\dag(T) \ket{f} }
\ee

\ni Now, the derivatives of the propagator only affect the corresponding sub propagator

\be
[\nabla U]_j = \frac{\p U}{ \p \ep_j} = U_M \cdots U_{j+1} \frac{\p U_j}{\p \ep_j} U_{j-1} \cdots U_1
\ee

\ni and the problem is reduced to the computation of

\be
\frac{\p U}{\p \ep_j} = -i\D \bar{H_c} U_j
\ee

\ni where

\be
\bar{H_c} = \frac{1}{\D} \int_0^{\D} U_j(-\t) H_c U_j(\t) d\t
\label{barH}
\ee

\ni That is, the gradient reads

\be
\begin{split}
\frac{\p I}{\p \ep_j} &=  -2 \D \Im[\bra{\chi_j} \bar{H}_c \ket{\psi_j}]\\
&=\int_{\D_j} g(t) dt \\
&=\int_0^{\infty} g(t) \sqcap_j(t) dt
\end{split}
\label{natu_integr}
\ee

\ni where in the last lines we have made explicit that the partial derivatives of the objective with respect to the parameters in the pulse are simply the projections of the so called \textit{natural} gradient \cite{natural,muller2015dressing}

\be
g(t) \equiv \frac{\delta I}{\delta \ep(t)} = -2 \Im[\bra{\chi(t)} H_c \ket{\psi(t)}]
\label{natgrad}
\ee

\ni on the basis functions of the PWC parametrization

\[
   \sqcap_j(t) =
    \begin{cases}
        1 & \text{if $t\in\D_j$}\\
        0 & \text{else}
    \end{cases}
\]

\ni Note that numerical integration of Eq. (\ref{natu_integr}) is a straightforward alternative for computing arbitrary precise approximations to the gradient. This option seems particularly suitable for K-GRAPE since, once the Krylov basis has been built for a given time step (and the effective Hamiltonian has been diagonalized), evaluating the evolved state on multiple points on a time-grid is virtually free. 

Another possibility for improving the quality of the gradient is to Taylor expand the exponentials in Eq. (\ref{barH}). A hierarchy of approximations to $\bar{H_c}$ unfolds

\be
\bar{H_c}^{(p)}=\frac{(-i\D)^p}{(p+1)!} \L^p H_c
\ee

\ni where we've introduced the Liouvillian operator $\L\equiv[H,\cdot]$ (Here, of course, the Hamiltonian corresponding to the corresponding time-slot should be used $H\equiv H_j$). This is ultimately translated into a hierarchy of approximations to the gradient

\be
\frac{\p I^{[P]}}{\p \ep_j} =  -2 \D \Im[\bra{\chi_j} \bar{H}^{[P]}_c \ket{\psi_j}]
\label{gradp}
\ee

\ni where the notation $[P]$ implies that we sum over $0\leq p \leq P$

\be
\bar{H_c}^{[P]} = \sum_{p=0}^{P} \bar{H_c}^{(p)}
\ee



A common practice for computing the matrix exponentials in Eq. (\ref{eq-matexp}) is to perform an eigendecomposition of the Hamiltonian and to invoke the spectral theorem

\be
U_j = \sum_{k=1}^D e^{-i \lambda_j \D_j} \ket{\lambda_k}\bra{\lambda_k} = Q D Q\dag
\ee

\ni here ${\ket{\lambda_k}}$ are the eigenvectors of $H_j$, $\lambda_k$ its eigenvalues and
\begin{equation}
 \begin{split}
    &Q\dag = [ \ket{\lambda_1}, \ldots, \ket{\lambda_D}]\\
    &D=diag(e^{-i\lambda_1\D}, \ldots, e^{-i\lambda_M \D})
\end{split}  
\end{equation}


\ni An alternative way of computing these matrix exponentials is using Padé approximations. Although slightly more expensive, the eigendecomposition has a lot more to offer, since it provides exact derivatives of the propagators
\be
\frac{\p U_j}{\p c_j} = Q \Tilde{H}_c F Q\dag
\ee

\ni where , $\Tilde{H}_c=Q\dag(-i H_c)Q$ and

\[
    [F]_{jk}=-i\D \bra{\lambda_j} H_j \ket{\lambda_k}
    \begin{cases}
        \omega_1 & \text{if $0<t<\Delta t$}\\
        \vdots \\
        \omega_M & \text{if $(M-1)\Delta t<t<T$}
    \end{cases}
\]

Finally, let us propose a simple way of improving the quality of the zeroth order approximation to the gradient (Eq. (\ref{gradI})). Consider the first component of the standard zeroth-order gradient 

\be
\frac{\p I}{\p \ep_1} \approx \frac{\p I^{(0)}}{\p \ep_1} =  -2 \D \Im[\bra{f} U_M \ldots  H_c U_1 \ket{i} \beta^*]     
\label{dId1}
\ee

\ni This is precisely $\D$ times the natural gradient of Eq. (\ref{natgrad}) evaluated at $t=\D$. By virtue of Eq. (\ref{natu_integr}) we know that the exact calculation involves the integration of the natural gradient in the whole time domain of the pulse (see Eq. (\ref{natu_integr})). We thus propose a centered version of the approximation, where the natural gradient is still assumed constant in the interval, but evaluated at $t=\D/2$, the center of the pulse
\be
\frac{\p I}{\p c_j} \approx \frac{1}{2}[\frac{\p I^{(0)}}{\p c_j} + \frac{\p I^{(0)}}{\p c_{j+1}}]
\ee

\ni We find this error to be $O(\D^3)$ instead of the characteristic $O(\D^2)$ of the standard zeroth order.





\bibliography{kpaper.bib}	
\end{document}